\documentclass[journal=jacsat,manuscript=article]{achemso}

\usepackage{chemformula} 
\usepackage[T1]{fontenc} 
\usepackage{bm,graphicx,subfigure}%
\usepackage{xcolor}

\author{Charles Lin}
\affiliation{Department of Electrical and Computer Engineering, University of Toronto, Ontario, Canada}
\author{Pohan Chang}
\affiliation{Department of Electrical and Computer Engineering, University of Toronto, Ontario, Canada}
\author{Yiwen Su}
\affiliation{Department of Electrical and Computer Engineering, University of Toronto, Ontario, Canada}
\author{Amr S Helmy}
\affiliation{Department of Electrical and Computer Engineering, University of Toronto, Ontario, Canada}
\email{a.helmy@utoronto.ca}

\title[]
  {Electrically-Reconfigurable Passive and Active Circuits in a Single Plasmonic Architecture}

\begin{document}



\begin{abstract}
  
Guided-wave plasmonic circuits are promising platforms for sensing, interconnection, and quantum applications in the sub-diffraction regime. Nonetheless, the loss-confinement trade-off remains a collective bottleneck for plasmonic-enhanced optical processes. Here, we report a unique plasmonic waveguide that can alleviate such trade-off and improve the efficiencies of plasmonic-based emission, light-matter-interaction, and detection simultaneously. Through different bias configurations, record experimental attributes such as normalized Purcell factor approaching $10^{4}$, 10-dB amplitude modulation with $<$1 dB insertion loss and fJ-level switching energy, and photodetection sensitivity and internal quantum efficiency of -54 dBm and 6.4 \% respectively can be realized within the same amorphous-based plasmonic structure. The ability to support multiple optoelectronic phenomena while providing performance gains over existing plasmonic and dielectric counterparts offers a clear path towards reconfigurable, monolithic plasmonic circuits.
\end{abstract}

Unrestricted by the diffraction limit, plasmonic structures can enable stronger modal confinement to enhance optical processes and miniaturize photonic circuits~\cite{Gramotnev:2010,Papaioannou:2011,DeLeon:2012}. However, practical utilization of plasmonic components is commonly hindered by the intrinsic metal absorption, which becomes increasingly severe with field confinement~\cite{Gramotnev:2010}. Consequently, plasmonic resonators often have a limited quality factor ($Q$) to effective mode volume ($V_{eff}$) ratio, otherwise known as the Purcell factor~\cite{Purcell:1946}, which impedes their potential use towards laser threshold reduction, exaltation of optical non-linearity, or enhancement of detection sensitivity~\cite{Vahala:2003,Wu:2006,Min:2009,Kauranen:2012}. Similarly, for non-resonant plasmonic components, the trade-off inevitably leads to higher noise level, significant insertion loss, as well as  device-specific waveguide configurations, the last of which further limits the footprint, speed, and fabrication tolerance at the circuit level~\cite{Muehlbrandt:2016,Ayata:2017,Heni:2019}. Even amongst rapid advancements in the field of plasmonics, a flexible waveguide architecture that can better capitalize on the light-matter interaction (LMI) offered by plasmonic modes, particularly in a guided-wave setting that can interface with planar integrated photonic circuits, is still elusive and highly sought-after.

In this work, we demonstrate a coupled hybrid plasmonic waveguide (CHPW) that can alleviate the loss-confinement trade-off (Figure 1a). Electrically, the CHPW is built by joining Schottky junction and metal-oxide-semiconductor (MOS) stack through a common metal layer~\cite{Su:2019}. Optically, it combines the advantages of coupled~\cite{Berini:2001} and hybrid~\cite{Alam:2007,Oulton:2008} plasmonic waveguide (HPW) modes to simultaneously achieve long-range propagation and nanoscale modal confinement. As depicted in Figure 1b, a low-loss supermode can be engineered though the coupling of surface plasmon polariton (SPP) and HPW modes. By tuning the thickness of the top dielectric layer, the field overlap within the dissipative metal region can be minimized through destructive interference (Figure 1c). As a result, the supermode propagation loss can be reduced to 0.02 dB/$\mu$m, orders of magnitude lower than the uncoupled SPP and HPW modes. Concurrently, it can be observed that modal energy is highly localized within the low-index dielectric layer, as the continuity condition of the displacement field needs to be satisfied~\cite{Alam:2007}. Thus, instead of redistributing the energy density into the non-metallic waveguide layers like existing long-range plasmonic structures, CHPW can maintain its modal confinement to 0.002 $\mu m^2$. Moreover, as loss remains $<$0.05 dB/$\mu$m and confinement changes by only 5 \% even if the top layer thickness deviates from optimal condition by 10\%, the design is highly fabrication tolerant and wavelength-insensitive. Recent experimental demonstration of CHPW has reported propagation loss of 0.03 dB/$\mu$m~\cite{Su:2019}. Detailed analysis of the CHPW's passive attributes can be found in Supplementary Section 1. 

Contrary to coupled-mode plasmonic structures that are designed with strictly enforced structural, material, and modal symmetries~\cite{Berini:2001,Chen:2012,Ma:2014}, CHPW can offer stronger modal tunability owing to its asymmetrical nature. As shown in Figure 1d, by simply increasing the waveguide width from 200 to 620 nm, the supermode propagation loss can be enhanced by two orders of magnitude, from 0.02 to $>$1 dB/$\mu$m at $\lambda$=1.55 $\mu$m. This is attributed to supmerode hybridization and field symmetry breaking, which only manifest if the two modes involved in the coupling have different dispersion characteristics (see Supplementary Section 1). As such, without any modification to its vertical cross-section, the CHPW nano-structure can be engineered to have either long-range propagation or strong dissipation over broad wavelength and temperature conditions.

The highly tunable characteristics of the CHPW architecture in turn enables previously unattainable functional versatility. First, in its low-loss state, CHPW can maintain a modest Q-factor without sacrificing $V_{eff}$. Recently, normalized CHPW Purcell factor up 6507 has been demonstrated~\cite{Su:2019}, making it an attractive platform for light generation. Second, by altering the modal interference condition through bias, the otherwise low-loss supermode can be efficiently induced into a highly-absorptive state. For example, if an indium tin oxide (ITO) layer is embedded into the MOS region, strong carrier dispersion effect can be induced via the epsilon-near-zero (ENZ) mechanism (Figure 1e)~\cite{Lin:2015}. As a result, amplitude modulation with high extinction ratio (ER) can be achieved while maintaining low insertion loss (IL), allowing CHPW to provide better utilization of the enhanced LMI afforded by plasmonics (see Supplementary Section 2). Finally, since the Schottky interface contained within the structure naturally supports internal photoemission (Figure 1f), the very same structure, in its widened high-loss state, can provide efficient photodetection. Recently, CHPW photodetector with sensitivity down to -35 dBm has been reported~\cite{Su:2017}. Such versatility is not observed in other waveguide platforms and can empower monolithic plasmonic circuit integration, as a single structure can provide either passive or active functionalities depending on biasing configurations (Figure 1g).

To illustrate the capability of the CHPW as a device platform, we fabricated and characterized CHPW micro-ring resonators, modulators, and photodetectors on a single chip. The devices are demonstrated using identical material stack (Si/Al/$\mathrm{SiO_2}$/Si) and layer thicknesses (185 nm/10 nm/20 nm/220 nm)  for the first time, with an 10 nm ITO layer added for the modulators. Note that ITO can also be included in other devices in the future if the as-deposited carrier density can be sufficiently low (see Supplementary Section 2). With exception of the Si-on-insulator substrate, all of the waveguide layers consisted of amorphous materials deposited through sputtering and evaporation (see Methods). Nonetheless, as will be shown below, CHPW can offer experimental performance gain for every device despite the monolithic nature and amorphous-based active layers. 

A scanning electron microscopy (SEM) image of a single-mode, 200 nm wide CHPW is shown in Figure~\ref{fig:ring}a. In comparison to reference HPWs fabricated on the same chip, the incorporation of an additional top Si layer leads to a 90 \% reduction in waveguide loss, from 0.59 to 0.07 dB/$\mathrm{\mu}$m at $\lambda$=1.55 $\mu$m (Figure~\ref{fig:ring}b). More importantly, from Figure~\ref{fig:ring}c, it is observed that our amorphous-based CHPW provides the best experimental loss-confinement balance compared to existing plasmonic waveguides. Hence, the adverse effect of Ohmic loss inherent to plasmonic structures is highly alleviated. In addition, the waveguide is shown to be suitable for photonic-plasmonic integration, as an end-fire coupling efficiency of 77 \% is measured between 750 nm wide Si nanowires and CHPWs. As observed in Supplementary Section 3, strong coupling can be sustained between $\lambda=$ 1450-1650 nm through momentum and mode matching.

The combination of minimal propagation loss and strong field localization in turn enables significant enhancement in the Purcell factor of CHPW micro-ring resonators (Figure~\ref{fig:ring}d). For an all-pass CHPW resonator with Si bus and radius of 2.05 $\mu$m, the loaded Q is measured to be 2,283 at room temperature and increases to 3,276 at 35$^{\circ}$C (Figure~\ref{fig:ring}e). To the author's knowledge, this is the highest experimental value reported for plasmonic micro-ring resonator to-date. More importantly, as displayed in Figure~\ref{fig:ring}f, the normalized Purcell factor is calculated to be 81,241, outperforming plasmonic and even dielectric counterparts. Furthermore, additional CHPW micro-ring characterization found in Supplementary Section 4 reveals that ER $>$ 10 dB can be maintained up to 50 $^{\circ}$C, which is an order of magnitude higher compared to ultrahigh-Q dielectric cavities. As a proof of concept, we have only manipulated the thickness of the top Si layer to optimize the CHPW supermode attributes. It is expected that further waveguide optimization or an extension to whispering-gallery disk structure will further enhance the resonator Purcell factor~\cite{Kwon:2012}.

Harnessing the strong absorption of widened CHPW nano-structure, sensitive photodetection have also been demonstrated (Figure~\ref{fig:pd}a). Specifically, micrometer-sized Al collector contacts are sputtered onto the intrinsic top Si layer of 620 nm wide CHPWs. As shown in Figure~\ref{fig:pd}b, the dark current noise ($I_{d}$) for a 15 $\mu$m photodetector increases with bias in accordance to the image-lowering effect. However, due to the compact device footprint, $I_{d}$ is always $<$0.1 $\mu$A and can be as low as $\sim$ 0.01 nA. At the same time, $\mu$A-level photocurrent ($I_p$) can be measured once the device is illuminated at $\lambda$= 1.55 $\mu$m. Such large current contrast is possible because (1) the use of amorphous junction relaxes the carrier momentum mismatch at the Schottky interface~\cite{Grajower:2018} and (2) the use of 10 nm thick Al emitter layer enhances the emission probability of hot carriers~\cite{Scales:2010b}. Therefore, responsivity of 82 mA/W and internal quantum efficiency (IQE) of 6.4 \% can be reached (Figure~\ref{fig:pd}c). At the Schottky reach-through voltage of 2 V, the continuous-wave sensitivity, which is the minimum incident optical power required to overcome the $I_d$ noise~\cite{Scales:2010b}, is only -54 dBm (Figure~\ref{fig:pd}d). 

In Supplementary Section 6, it is shown that both the CHPW sensitivity and IQE are the highest for plasmonic Schottky photodetectors reported to-date, with the latter approaching the performance previously only attainable through graphene-enhanced internal photoemission~\cite{Goykhman:2016}. Moreover, through the testing of a 5-$\mu$m CHPW photodetector, it is observed that sensitivity can be maintained between $\lambda$ = 1.53-1.57 $\mu$m and temperatures up to 100 $^{\circ}$C (Figure~\ref{fig:pd}d). Using available instrumentation, the speed of the photodetector is confirmed up to 26 GHz (Figure~\ref{fig:pd}e). However, with experimental parasitic capacitance of 12 fF, the RC response of the fabricated CHPW photodetectors have a cut-off frequency of 265 GHz (see Supplementary Section 5). 

The ability for CHPW to better capitalize on the plasmonically enhanced LMI for active functionalities is demonstrated through CHPW modulators as shown in Figure~\ref{fig:mod}a. With 185 nm top Si thickness, the IL of a 10 $\mu$m long CHPW with an embedded ITO layer is only $\sim$1 dB from $\lambda$= 1.52-1.6 $\mu$m (Figure~\ref{fig:mod}b). However, once a forward bias is applied to the Al-$SiO_{2}$-ITO stack and sufficient carrier accumulation is induced inside the ITO layer (beyond 20 V), optical transmission starts to decrease rapidly and ER $>$10 dB can be reached (Figure~\ref{fig:mod}c). Such high ER does not degrade up to 100 $^{\circ}$C  (see Supplementary Section 7) and is attributed to the CHPW's strong field localization, which enhances the effect of the ENZ mechanism~\cite{Lin:2015}. Note that no change in transmission is observed under the reverse bias condition or in reference structure without the ITO layer. Thus, the measured modulation response can be attributed to field-induced carrier-accumulation instead of drift in the set-up, thermal-optic effect at high bias voltage, or electro-optic effect in materials other than ITO.

Note that the energy-efficiency of these proof-of-concept CHPW modulators are only limited by the properties of the MOS oxide layer. By replacing the 20 nm sputtered $\mathrm{SiO_2}$ with 5 nm sputtered $\mathrm{TiO_2}$, the static voltage requirement is reduced by 50 \% (Figure~\ref{fig:mod}c). It is expected that the use of high-$\kappa$ oxide, such as $\mathrm{HfO_2}$, deposited via atomic layer deposition will enable CMOS-compatible driving voltage~\cite{Wood:2018}. Moreover, in the high bias regime, dynamic voltages of only 1.5 V and 3.1 V are sufficient to achieve 3-dB and 6-dB amplitude modulation respectively (Figure~\ref{fig:mod}d), corresponding to switching energy of a few fJ. Finally, as shown in Figure~\ref{fig:mod}e, the modulator frequency response is also flat up to 26 GHz and the RC cut-off frequency is $\sim$636 GHz (see Supplementary Section 8). A detailed comparison between experimental ITO-based plasmonic modulators can be found in Supplementary Section 9. Overall, CHPW provides the lowest IL as well as the highest optical figure-of-merit (ER/IL) and speed to-date. 

In conclusion, we report the first experimental demonstration of a plasmonic waveguide architecture that not only alleviates the plasmonic loss-confinement trade-off, but is also capable of simultaneously utilizing multiple device physics to enable electrically-driven functionalities using a single structure. Based on an unique class of asymmetrically-coupled supermode, our amorphous-based CHPW devices demonstrated record Purcell factor for plasmonic micro-rings, record quantum efficiency and sensitivity for guided-wave Schottky detectors at $\lambda=1.55 \mu$m, as well as record ER-IL ratio for plasmonic modulators based on ENZ effect. The ability to be simultaneously optimized for multiple passive and active functionalities is not observed in other plasmonic or dielectric waveguide platforms. Such demonstration provides a direct path towards the realization of monolithic plasmonic circuitry, one that has reconfigurable functional elements, experiences smaller inter-component coupling loss, and enables reduced fabrication complexity and cost~\cite{Ayata:2017,Krasavin:2011,Liu:2016}. 
While the CHPW devices demonstrated here are immediately applicable for on-chip optical interconnection, they also usher in new paths to many other potential applications. Specifically, the ultra-high Purcell factor can either help reduce thermal effects for biosensing and optical trapping~\cite{Anker:2009} or increase the rates of spontaneous and stimulated emission to reduce threshold gain of nanolasers~\cite{Wei:2016}. By coupling to quantum emitters, CHPW resonators can also be applied towards long-range quantum entanglement for the implementation of single-photon transistors~\cite{Chang:2007}. Finally, the strong modal field localization can be used to improve the quantum efficiency of single photon detector~\cite{Eftekharian:2013} while the ENZ effect can be applied towards ultrafast all-optical modulation~\cite{Alam:2016}.

{\large \textbf{Methods}}\\ \\
\textbf{CHPW device fabrication}: The CHPW devices were fabricated on SOITEC wafer with 220 nm thick crystalline Si layer. To create the multi-layer waveguide stack, 10 nm thick ITO and 200 nm thick top Si layers were deposited using AJA International ATC Orion 8 Sputter Deposition System in Ar plasma. An undoped Si target was used with bulk resistivity $> 1 \Omega-cm$.The 20 nm thick $\mathrm{SiO_2}$ and 10 nm thick Al were deposited using Angstrom Nexdep Electron Beam Evaporator. The root-mean-square roughness of the deposited stack was measured using atomic force microscopy to be 1.49 nm. Multiple lithography steps were required to define the CHPW devices and they were carried out using the Vistec EBPG 5000+ Electron Beam Lithography System. For device patterning, the Si and $\mathrm{SiO_2}$ layers were etched using the Oxford Instruments PlasmaPro Estrelas 100 DRIE System with a $\mathrm{SF_6}$:$\mathrm{C_4F_8}$ mixed-gas plasma while the ITO layer was etched using a gas mixture of $\mathrm{Cl_2}$/$\mathrm{BCl_3}$/Ar via the Trion Minilock II ICP-RIE System. Electrical contacts for the CHPW modulators and photodetectors were fabricated by sputtering 200 nm thick Al onto the devices. The experimental device dimensions were confirmed with scanning electron microscopy.\\ \\

\textbf{ITO Sputtering}: ITO was deposited using AJA International ATC Orion 8 Sputter Deposition System at pressure of 3 mTorr. Gas combination during sputtering was 80:1 Ar:$\mathrm{O_2}$ and power was 50 W. The electronic properties of the ITO were characterized by four-point probe and Hall measurements. \\ \\

\textbf{CHPW device characterization}: To perform cut-back measurement, light from a C-band continuous-wave laser was first amplified using an Erbium Doped Fiber Amplifier (EDFA). Next, the output fiber from the EDFA was wrapped through a Thorlabs paddle fiber polarization controller to ensure TM-polarized input. A single-mode lensed fiber that has a 2.5 $\mu$m spot diameter was used to couple light into the input Si nanowires. On the output side, the transmitted light from the output Si nanowire was collected with a 20x objective lens with 0.4 numerical aperture. Noise from the substrate was eliminated using an iris and the output polarization was confirmed with a polarization beam splitter. At last, germanium photodetector was used for power measurement.

Thermal measurement was performed using a custom copper stage with thermoelectric coolers, where temperature was controlled using the Keithley 2510-AT Autotuning TEC source meter through electrical feedback from a 10k$\Omega$ thermistor.

To measure the optoelectronic bandwidth of the CHPW modulators up to 6 GHz, RF signal was generated by the NI PXIe-5673E RF vector signal generator (VSG) with 6 dBm peak-to-peak amplitude and a rectangular shape. A bias-tee set before the input of the RF probe was used to combine the RF signal with DC offset from a Keithley 2604B source meter. The resulting amplitude modulated optical signal was captured via a New Focus 1454 18.5 ps Vis-IR photodetector and amplified with an AMF 00101000-55 amplifier before being fed into the NI PXIe-5663E vector signal analyzer (VSA) for analysis. Similarly, the electrical bandwidth of the CHPW photodetector was quantified using the NI VSG and VSA. On the input side, the VSG stimulus was amplified and drives a Fujitsu $\mathrm{LiNbO_3}$ Mach-Zehnder modulator, generating intensity-modulated optical signals up to 6 GHz. On the output side, the photocurent was collected with a Cascade Infinity probe and amplified before feeding into the VSA. In order to subtract coherent electrical input-output crosstalk, measurements with laser beam turned off were also performed. For measurement between 6 GHz and 26 GHz, the same set-up with Keysight M9375A 26.5 GHz 6-port Vector Network Analyzer and AMF-4F-060180 amplifier were used.


\begin{suppinfo}

The following files are available free of charge.
\begin{itemize}
  \item Additional analysis and description on the design and characterization of coupled hybrid plasmonic waveguide, resonator, modulator, and photodetector.
\end{itemize}

\end{suppinfo}


\bibliographystyle{Science}

\begin{figure*}
	\centering
	\includegraphics[width=0.9\linewidth]{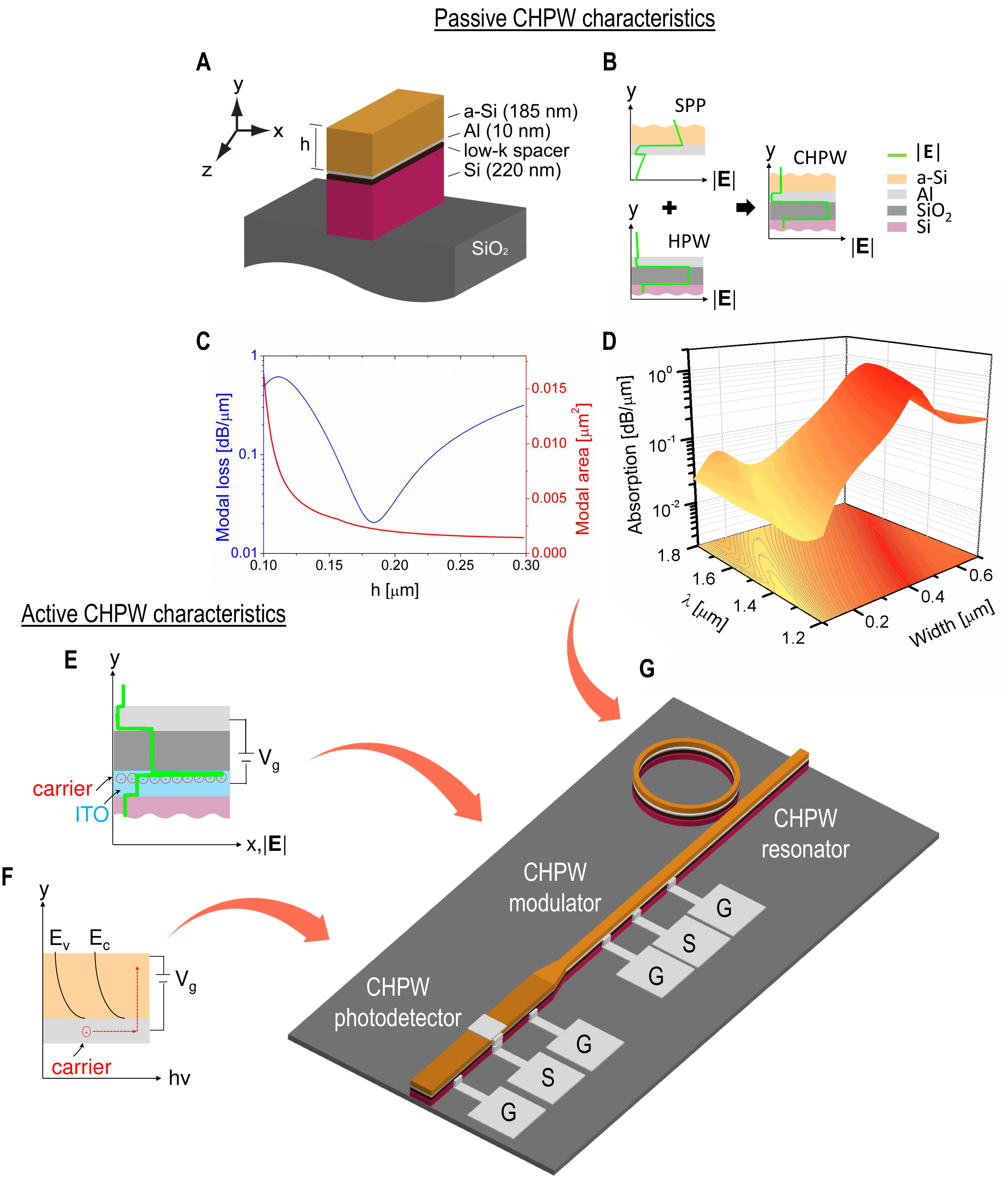}
	\caption{\textbf{CHPW Device Platform:} (\textbf{A}) Schematic of the multilayer CHPW structure. (\textbf{B}) 1D field profiles of the modes supported by the CHPW structure. A long-range, transverse-magnetic supermode can be engineered through the out-of-phase coupling of SPP and HPW modes to reduce the field-overlap in the metal region. Note that a highly-lossy supermode that corresponds to constructive modal interference also exists (see Supplementary Section 1). (\textbf{C}) Calculated modal area and propagation loss of the low-loss CHPW supermode as function of the top Si layer thickness ($h$). (\textbf{D}) Tunability of the long-range CHPW supermode as function of waveguide width and operation wavelength. (\textbf{E}) CHPW-based photodetection based on internal photo-emission at the Schottky interface~\cite{Su:2017}. (\textbf{F}) CHPW-based amplitude modulation using ENZ effect in an embedded ITO layer~\cite{Lin:2015}. (\textbf{G}) Vision of monolithic CHPW plasmonic circuitry.}
	\label{fig:platform}
\end{figure*}

\begin{figure*}
	\centering
	\includegraphics[width=1.0\linewidth]{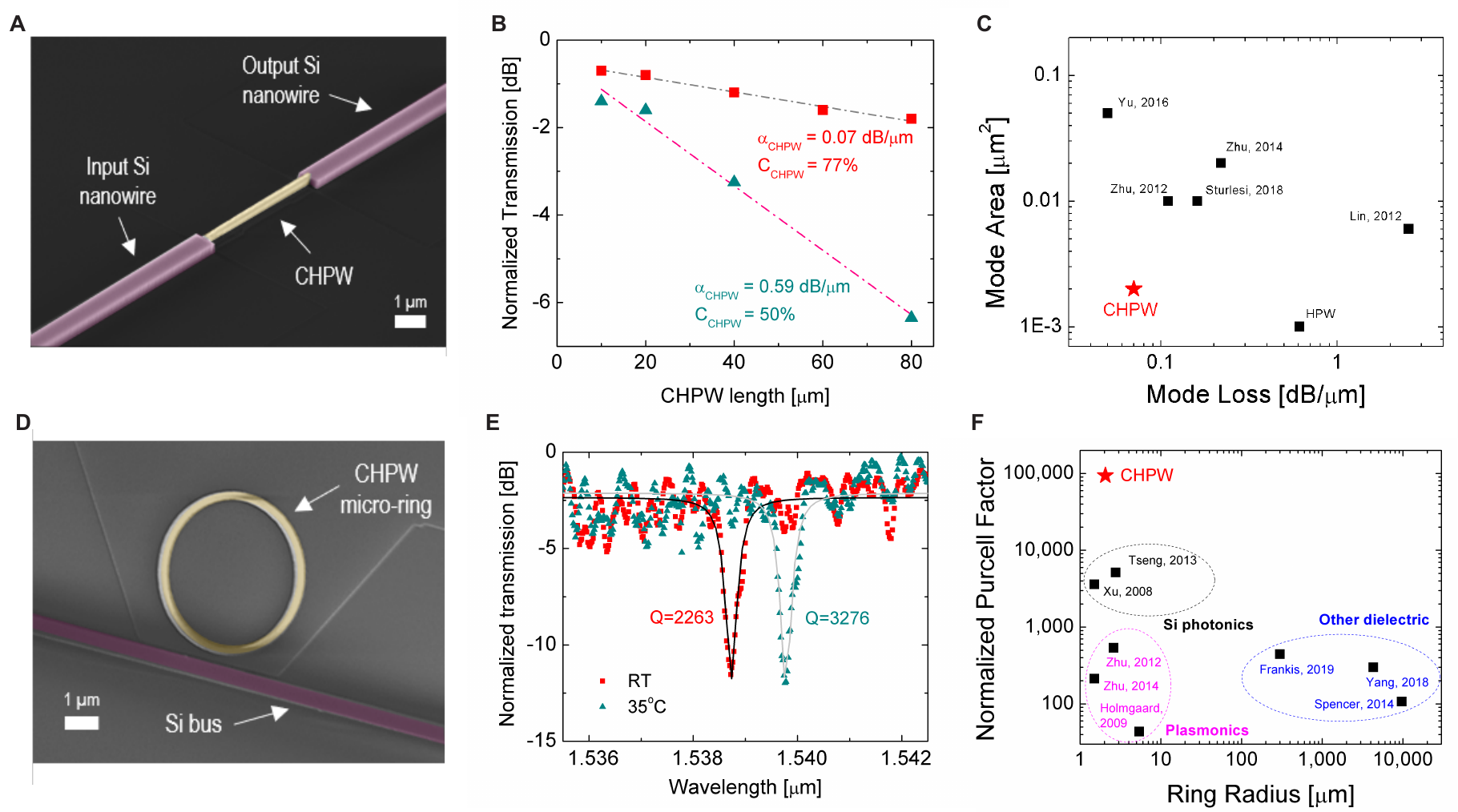}
	\caption{\textbf{Characterization of Passive CHPW Components:} (\textbf{A}) False-colored SEM image of a 5 $\mathrm{\mu}$m long CHPW. (\textbf{B}) CHPW cut-back measurement result. Data for reference HPW (no top $\alpha$-Si layer) is also plotted. (\textbf{C}) Loss-confinement comparison between experimentally demonstrated plasmonic waveguides. (\textbf{D}) Colorized SEM image of CHPW ring resonator, with 2.05 $\mu$m radius, 215 nm gap width, and 450 nm Si bus. (\textbf{E}) Measured transmission spectra of CHPW micro-ring, fitted to Lorentzian functions. (\textbf{F}) Normalized Purcell factor comparison between experimentally demonstrated micro-ring resonators. References: Lin, \textit{Appl. Phys. Lett.} \textbf{101,} 123115 (2012); Zhu, \textit{Opt. Express} \textbf{20,} 15232-15246 (2012); Zhu, \textit{IEEE PHOTONIC TECH L.} \textbf{26,} 833-836 (2014); Yu, \textit{Nano Lett.} \textbf{16,} 362-366 (2016); Sturlesi, \textit{APL Photonics} \textbf{3,} 036103 (2018); Xu. \textit{Opt. Express} \textbf{16,} 4309-4315 (2008); Tseng, \textit{Opt. Express} \textbf{21,} 7250-7257 (2013); Holmgaard, \textit{Appl. Phys. Lett.} \textbf{94,} 051111 (2009); Spencer, \textit{Optica} \textbf{1,} 153-157 (2014); Yang, \textit{Nat. Photonics} \textbf{12,} 297-302 (2018); Frankis, \textit{Opt. Lett.} \textbf{44,} 118-121 (2019).}
	\label{fig:ring}
\end{figure*}

\begin{figure*}
	\centering
	\includegraphics[width=0.8\linewidth]{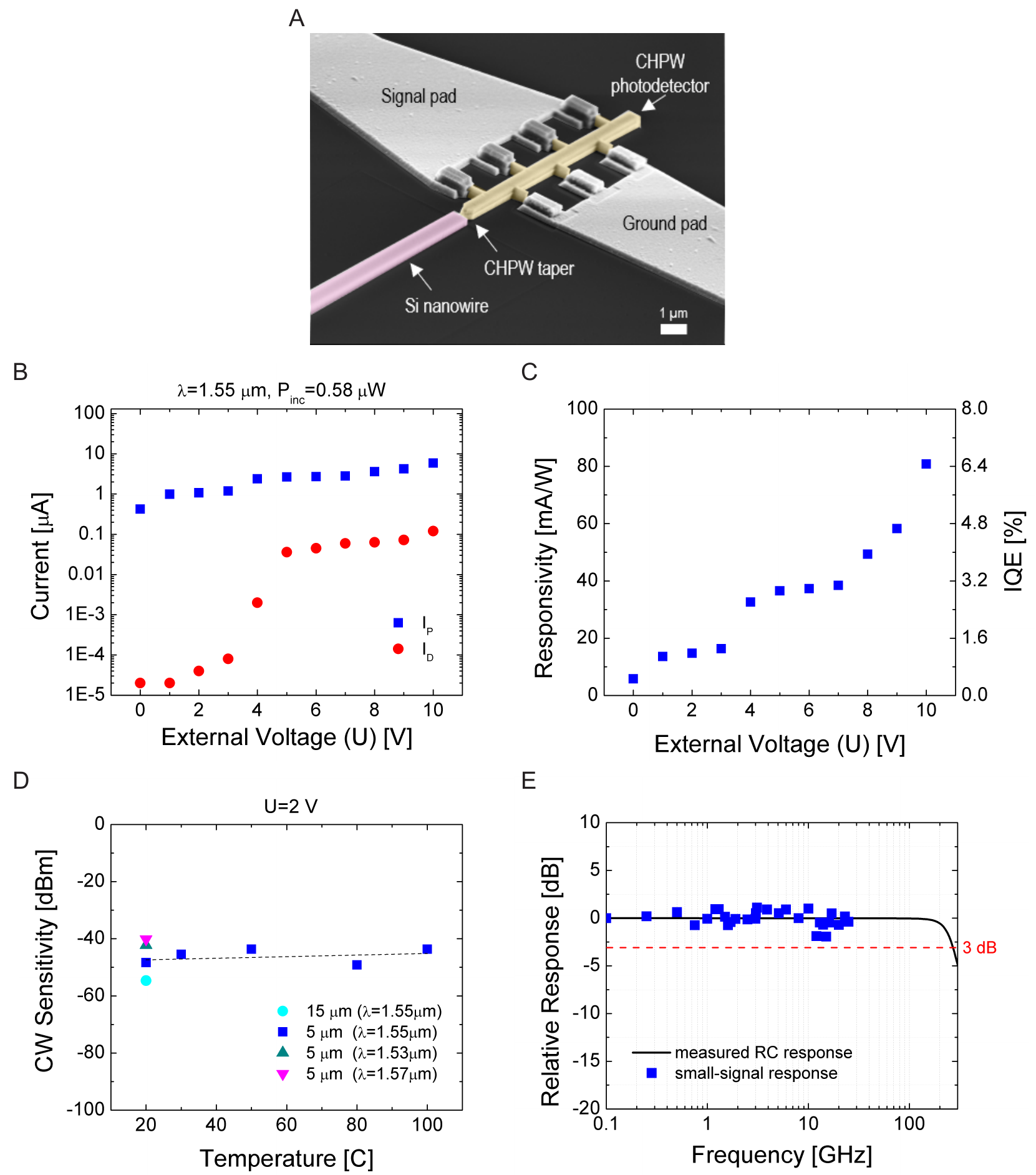}
	\caption{\textbf{Characterization of CHPW Photodetectors:} (\textbf{A}) False-colored SEM image of a 15 $\mathrm{\mu}$m CHPW photodetector. To minimize disturbance to the optical mode, micrometer-sized contacts are made to the Al and $\alpha$-Si layers using finger structures. Light is first coupled into a 200 nm wide CHPW, which then tapers over 500 nm to a width of 620 nm. (\textbf{B}) Detector photocurrent ($I_p$) and dark current ($I_d$) at $\lambda$=1.55 $\mu$m. (\textbf{C}) Responsivity and internal quantum efficiency at room temperature and $\lambda$=1.55 $\mu$m. (\textbf{D}) Temperature dependence of the detector's continuous-wave sensitivity for different device lengths and operation wavelengths. The devices are tested up to equipment limitation of 1.57 $\mu$m but is expected to be operational up to 1.8 $\mu$m based on simulation. (\textbf{E}) Frequency response of a 5 $\mathrm{\mu}$m CHPW photodetector measured up to 26 GHz and fitted to its RC response, modeled using a 50 $\Omega$ load resistance and 12 fF measured capacitance. The measurement is normalized to the response obtained at 100 MHz. }
	\label{fig:pd}
\end{figure*}

\begin{figure*}
	\centering
	\includegraphics[width=0.8\linewidth]{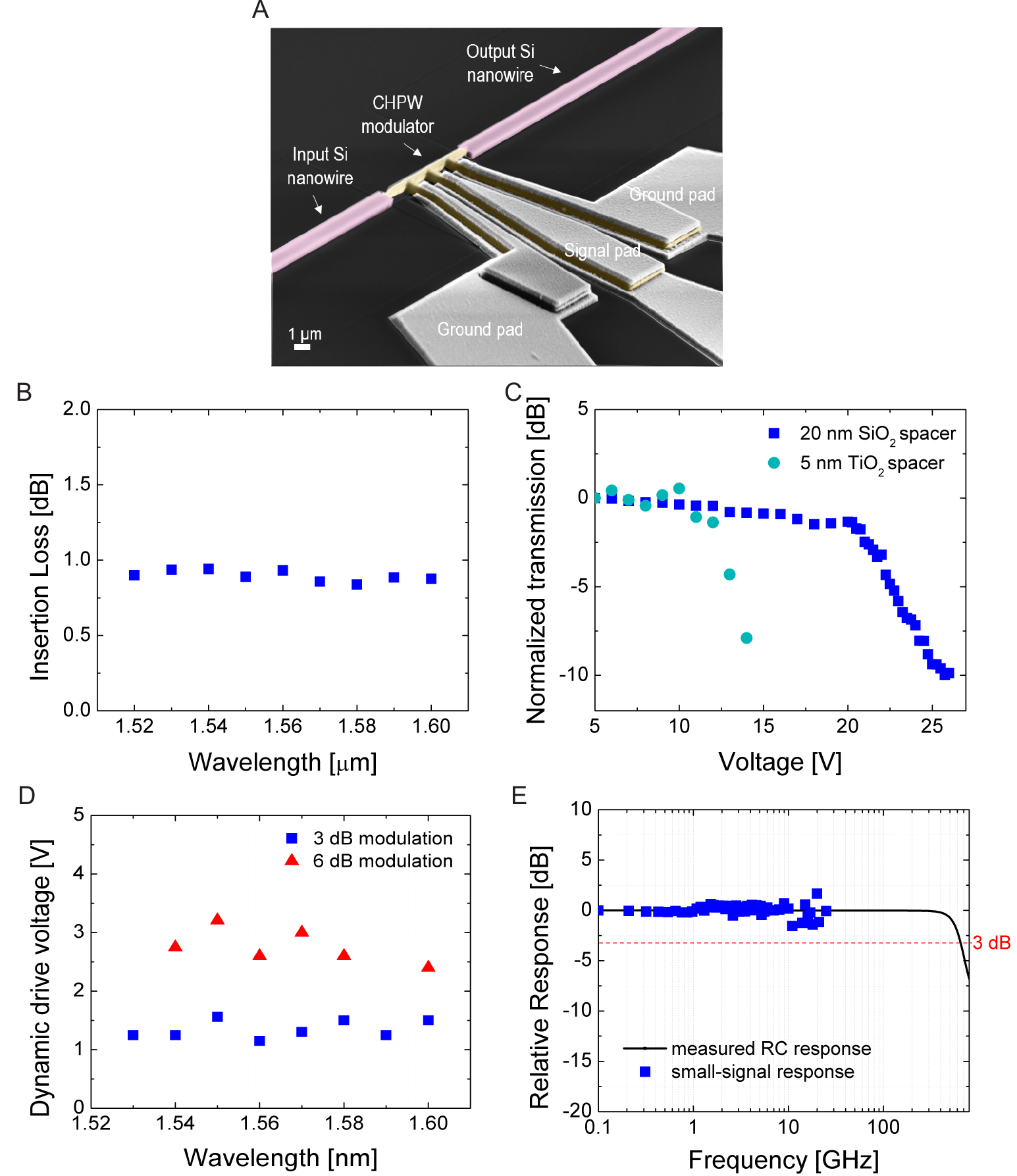}
	\caption{\textbf{Characterization of CHPW Modulators:} (\textbf{A}) False-colored SEM image of a 10 $\mathrm{\mu}$m long CHPW modulator. Under an external bias, carriers are injected through the contacts and the formation of voltage-induced accumulation layer at the ITO-$\mathrm{SiO_2}$ interface then enables optical modulation through ENZ effect. (\textbf{B}) Insertion loss of the CHPW modulator, showing wavelength-insensitivity due to the non-resonant nature of the CHPW supermode. (\textbf{C}) Transfer function of the CHPW modulator with different oxide materials and thicknesses. (\textbf{D}) Measured dynamic drive voltage requirement necessary for 3-dB and 6-dB amplitude modulation using a 20 nm $\mathrm{SiO_2}$ spacer layer. (\textbf{E}) Optoelectronic bandwidth of the modulator. The measurement (including the effects from RF probe and substrate characteristics) is normalized to the response obtained at 100 MHz. }
	\label{fig:mod}
\end{figure*}

\end{document}